\documentclass[aps,prd,twocolumn,superscriptaddress,nofootinbib]{revtex4}

\usepackage{graphicx}

\usepackage{amsmath}

\usepackage{amssymb}

\usepackage{pstricks}

\begin{document}

\title{Quantisation of 
Parameters and the
String Landscape Problem}~\thanks{This research work was
supported by the grants POCI/FP/63916/2005,
FEDER-POCI/P/FIS/57547/2004 and Ac\c c\~oes Integradas
(CRUP-CSIC) Luso-Espanholas E-138/04}

\author{Mariam Bouhmadi-L\'{o}pez}
\email{mariam.bouhmadi@fisica.ist.utl.pt}
\affiliation{Centro Multidisciplinar de Astrof\'{\i}sica - CENTRA, Departamento de F\'{\i}sica, Instituto Superior T\'ecnico, Av. Rovisco Pais 1,
1049-001 Lisboa, Portugal}
\affiliation{Departamento de F\'{\i}sica,
Universidade da Beira Interior, Rua Marqu\^{e}s d'Avila e Bolama,
6201-001 Covilh\~{a}, Portugal}
\affiliation{Institute of
Cosmology and Gravitation, University of Portsmouth,  Mercantile
House, Hampshire Terrace,  Portsmouth  PO1 2EG, UK}

\author{Paulo Vargas Moniz}
\email{pmoniz@ubi.pt}
\affiliation{Departamento de F\'{\i}sica,
Universidade da Beira Interior, Rua Marqu\^{e}s d'Avila e Bolama,
6201-001 Covilh\~{a}, Portugal} 
\affiliation{Centro Multidisciplinar de Astrof\'{\i}sica - CENTRA, Departamento de F\'{\i}sica, Instituto Superior T\'ecnico, Av. Rovisco Pais 1,
1049-001 Lisboa, Portugal}

\begin{abstract}
We broaden the domain of application of 
Brustein and de Alwis
recent paper \cite{Brustein:2005yn},
 where they introduce a (dynamical) 
selection principle on the landscape of string solutions 
using FRW quantum cosmology. More precisely, we (i) explain 
how their analysis is based in choosing a restrictive range of parameters,
thereby affecting the validity of the predictions extracted and (ii)  
subsequently  provide a wider and cohesive description,  regarding the 
probability distribution induced by quantum cosmological transition amplitudes. 
In addition, employing  DeWitt's argument \cite{DeWitt1}
for an initial condition on the wave function of the Universe,
we found that the string and
gravitational  parameters
become related through interesting  expressions involving an integer $n$,
suggesting a quantisation relation for some of the involved
parameters.
\end{abstract}

\date{\today}

\maketitle

\section{Introduction}\label{sect1}

The existence of a  multiverse of vast solutions 
\cite{Bousso:2000xa,Douglas:2003um,Susskind:2003kw} 
to string theory \cite{Polchinski2:1998rr} constitutes 
currently an important challenge \cite{Polchinski:2006gy}: 
How to  select a Universe or a class from the multiverse  
that will bear significant similarities  to ours?

The framework of quantum cosmology 
\cite{Vilenkin:1983xq,Hartle:1983ai,Linde:1983cm} 
provides a methodology  to establish a probability distribution for the 
dynamical parameters of the universe. This distribution probability  depends 
on the boundary condition chosen for the wave function of 
the Universe \cite{Hawking:1984hk,Vilenkin:1984wp, Linde}. 
Extending the procedure towards string theory requires  the 
inclusion of the dynamics of the moduli (namely, their 
eventual stabilisation through, e.g., fluxes and 
non-perturbative effects \cite{Giddings:2001yu,Kachru:2003aw,Douglas:2006es}). 
A simple setting where this quantum cosmological approach can be tested 
is that for which 
the moduli, the dilaton $S$, and the volume modulus 
$T$ are fixed by stringy effects (cf. ref. \cite{Brustein:2005yn}).

Recently,  
R.~Brustein and S.~P.~de Alwis (BA)  proposed in \cite{Brustein:2005yn},
using FRW quantum cosmology,  a  
dynamical selection principle  on the 
landscape\footnote{However, see references 
\cite{Banks:2004xh,Banks:2003es} for an  sceptical point of view  
about the string landscape issue.} of string solutions \cite{Susskind:2003kw}, 
without reference to the anthropic principle  \cite{Vilenkin:2004fj,Weinberg:2005fh}.
In more detail, this selection principle employs a
 thermal boundary condition for the wave function of the Universe 
and leads  to results similar to those found by  Sarangi 
and Tye\footnote{See also references 
\cite{Firouzjahi:2004mx,Huang:2005wq,Sarangi:2006eb,Barvinsky:2006uh,Holman:2005eu,Watson:2006px}.} 
\cite{Sarangi:2005cs}. This boundary condition states that  
the Universe emerges from the string era in a thermally 
excited state above the Hartle-Hawking (HH) vacuum, allowing to 
determine a tunnelling probability to different points in the landscape. 
Furthermore, this primordial thermal bath is   effectively described 
by a radiation fluid  which ``depends'' on the cosmological constant. 
Moreover, these ans\"{a}tze imply a switch   in the usual features of the HH \cite{Hartle:1983ai} 
and the tunnelling (Vilenkin) \cite{Vilenkin:1983xq,Vilenkin:1998rp} 
(see also \cite{Linde:1983cm,Linde} for a variation) wave functions:
The  HH wave function,  once the  thermal boundary condition of \cite{Brustein:2005yn}
is assumed, favours a non-vanishing cosmological constant larger than 
the one preferred by the tunnelling (Vilenkin) wave function. The extension 
to a model  with {\em dynamical} moduli was also presented in 
\cite{Brustein:2005yn}, applying it, e.g.,  to the model of Kachru, 
Kallosh, Linde and Trivedi (KKLT) \cite{Kachru:2003aw}. BA 
subsequently selected a class of values for  the parameters that 
characterise a  multiverse in the landscape \cite{Brustein:2005yn}. 

Such a quantum cosmological scenario 
constitutes indeed an interesting and 
promising framework to address the string landscape 
problem\footnote{It should be noticed that there is an 
alternative methodology to investigate the multiverse 
in the string landscape. It is based on the statistics of 
the solutions \cite{Douglas:2003um,douglas1,douglas2,douglas3} and 
claims that the number density of solutions is uniform 
as a function of the value of the cosmological constant.}. 
Nevertheless, there are a few significant points that do 
require additional investigation
and which we address in this paper as follows.

In Section \ref{sect2} we will review critically 
the issue of determining transition tunnelling amplitudes for a closed 
FRW Universe with a positive cosmological constant, $\Lambda$, and a 
radiation fluid (i.e., the simpler background introduced by BA in 
\cite{Brustein:2005yn}; Note that  a detailed description of a FRW 
minisuperspace with a
cosmological constant and a radiation fluid can be found in
\cite{Halliwell:1989ky,Bouhmadi-Lopez:2002qz}).
In Section \ref{sect3}, we thoroughly analyse the thermal boundary 
condition proposed in \cite{Brustein:2005yn},  clarifying the 
approximations used therein and the limits of their applicability. 
In particular, we discuss how making the amount of radiation  to ``depend" 
in a very particular way on the cosmological constant is the key of  BA  
new insight (see Eq.~(\ref{newro})), thereby switching the roles  of the 
Hartle-Hawking \cite{Hartle:1983ai} and 
tunnelling (Vilenkin) \cite{Vilenkin:1983xq,Vilenkin:1998rp} 
 wave functions in the sense we have previously mentioned.
Moreover, we prove that the thermal boundary condition applied in  
\cite{Brustein:2005yn} corresponds to the particular physical situation 
where the amount of radiation is very large.
Section \ref{sect4} constitutes an important element of our analysis. 
Therein, we explain how the BA analysis is based on a  narrow
approximation range and not fully sustainable. We then provide a broader
and improved 
analysis that is independent of such  restrictive limit; i.e. we consider 
an arbitrary amount of radiation consistent with the tunnelling of a 
radiation-filled Universe with a positive $\Lambda$.
Furthermore, we point out 
that the switching of the roles of the Hartle-Hawking \cite{Hartle:1983ai}  
and tunnelling (Vilenkin) \cite{Vilenkin:1983xq} 
wave functions is not mandatory, when investigating
the probability of tunnelling to different points in the landscape,
under the presence of  a radiation term which depends 
on the cosmological constant.

The FRW minisuperspace employed in the previous 
Sections allows to extract additional new results that 
were not the focus of interest in 
\cite{Brustein:2005yn}. In fact, in Section \ref{sect5},  
we  explain how a  quantum mechanical formulation allows 
to determine
how some  parameters, which may characterise a multiverse 
in the landscape, would be related through  an integer. 
This result is obtained by imposing  DeWitt's   condition 
\cite{DeWitt1,Davidson:1999fb,Bouhmadi-Lopez:2004mp}; i.e. the wave function  vanishes at a classical singularity,
to address the presence of a divergence in the
curvature.
More precisely,  the employed FRW minisuperspace (cf.~\cite{Brustein:2005yn}) 
has such a divergence in a classically
allowed region. This singularity can be dealt with, not
by  replacing it with an Euclidean conic-singularity-free pole, but instead,
 by making it quantum mechanically
 inaccessible \cite{DeWitt1,Davidson:1999fb,Bouhmadi-Lopez:2004mp},
therefore turning  it  neutralised.
In Section \ref{sect6},  we include moduli fields 
similarly to \cite{Brustein:2005yn}, but  adapting  in a consistent
manner  the improved thermal boundary condition  treatment of 
Section 4 towards a KKLT-like setting. 
Finally, in Section \ref{sect7} we present a summary of our work,
discussing the physical implications of our results.

For completeness, let us point out the relation between the notation used in \cite{Brustein:2005yn}
and the followed in this paper ($\sigma^2=2\textrm{G}/(3\pi)$): 

\begin{center}

\begin{tabular}{|c|c|}
\hline  Our notation & BA notation \\
\hline  $\lambda$ &  $\sigma^{-2}\lambda$ \\
\hline  $\tilde{K}$ & $\sigma^{2}K$ \\
\hline  $\nu$  & $\sigma^{-2}\nu$ \\
\hline  $a^2$ & $\sigma^{2}a^2$ \\
\hline  $t$ time & $\sigma t$ \\
\hline $\rho$ & $\frac{1}{4\pi^2}\rho$ \\
\hline  $C$ & $\frac{1}{4\pi^2}C$ \\
\hline  $b^4$ & $4\pi^2 c^4$ \\
\hline  $I$ & $\Phi$ \\
\hline
\end{tabular}

\end{center}


\section{FRW Transition Amplitudes with 
Radiation and a positive Cosmological Constant}\label{sect2}

The aim of this section is
to review 
(cf. Refs.~\cite{Vilenkin:1998rp,Halliwell:1989ky,Bouhmadi-Lopez:2002qz,Rubakov:1984bh}) 
the standard framework 
concerning transition amplitudes for 
 a closed
radiation-filled FRW Universe with a cosmological constant 
and for two well-known boundary conditions: 
the 
Hartle-Hawking (HH) \cite{Hartle:1983ai} and 
tunnelling (Vilenkin) \cite{Vilenkin:1983xq}. 

\begin{figure}[h]
\begin{center}
\includegraphics[width=0.9\columnwidth]{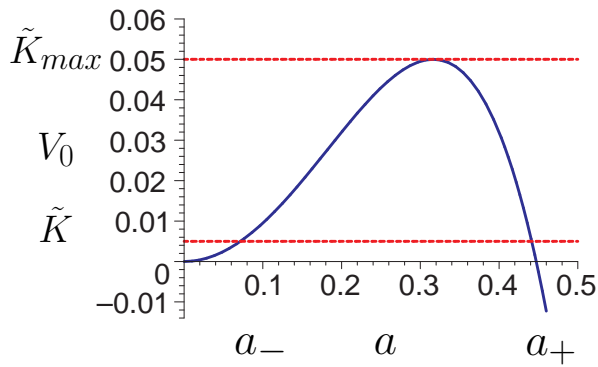}
\end{center}
\caption{The blue solid line represents the 
potential $V_0$ defined as $V(a)$ given in Eq.~(\ref{WdW})
in absence of radiation; i.e. for $\tilde{K}=0$. The red 
dashed-lines represent different amounts of radiation given by
$\tilde{K}$ and $\tilde{K}_{max}$. The turning points, $a_{\pm}$, 
are  the solutions  of $V_0(a)=\tilde{K}$.
The larger is the amount of radiation; i.e. the larger is $\tilde{K}$, 
the closer are the turning points. For 
$\tilde{K}=\tilde{K}_{max}$, $a_+$ and $a_-$ coincide. This amount 
of radiation represents the border line between having a tunnelling 
effect and no tunnelling effect. We will refer to a Universe with 
an amount of radiation such that 
$\tilde{K}$ is close to $\tilde{K}_{max}$ (but still smaller 
than $\tilde{K}_{max}$) as a Universe filled with a 
large amount of radiation.}\label{p9}
\end{figure}

The Wheeler-DeWitt equation for a closed FRW Universe filled with
radiation, represented by a perfect fluid whose energy density is given 
by $\rho=3\tilde{K}/(8\pi\textrm{G}a^4) $, and a positive
cosmological constant, $\Lambda$ ($\Lambda\equiv 3\lambda$), reads
\begin{eqnarray}\label{WdW}
\left[-\frac{\textrm{G}}{3\pi}\frac{d^2}{d\,
a^2} + \frac{3\pi}{4\textrm{G}} V(a)\right]\Psi (a)=0,&&\\
V(a)=a^2-\lambda a^4 -\tilde{K}.&& \nonumber 
\end{eqnarray} 
In the previous equations $a$ is the scale factor, $\textrm{G}$ is 
the gravitational constant and $\tilde{K}$ is a 
parameter that measures the amount of radiation. It can be seen 
that there are two turning points $a_{\pm}$ for $0<4\tilde{K}\lambda<1$, 
where the potential $V$ vanishes (see Fig.~\ref{p9}): 
\begin{equation}
a^2_{\pm}=\frac{1\pm{m}}{2\lambda},\quad
m=\sqrt{1-4\tilde{K}\lambda}.
\label{eq2}\end{equation}
These turning points split the Lorentzian and Euclidean regions. 
We also note that the maximum of the potential $V$ is reached at 
\begin{equation}
a_{max} \equiv a_* = \sqrt{1/2\lambda},
 \label{n2}
\end{equation}
where
\begin{equation}
V(a_*)= \frac{1}{4\lambda} - \tilde{K}.
 \label{n3}
\end{equation}
Therefore, there will be always an Euclidean region as long as the condition
\begin{equation}
V(a_*)>0 \Leftrightarrow 4\lambda\tilde{K}<1
 \label{n4}
\end{equation}
holds. Consequently, the maximum amount of radiation consistent with the tunnelling of the Universe  is such that $\tilde{K}<\tilde{K}_{max}$ (see Fig.~\ref{p9}), where 
\begin{equation}
\tilde{K}_{max}\equiv\frac{1}{4\lambda}.
\end{equation}
From now on, for a given positive cosmological constant, we will refer to a Universe as filled with a large amount of radiation if the radiation energy density is such that $\tilde{K}$ is close to $\tilde{K}_{max}$ but still smaller than $\tilde{K}_{max}$ (see Fig.~\ref{p9}).

The transition amplitude, $\mathcal{A}$,
of the Universe to evolve from the first Lorentzian region ($a<a_-$) to 
the larger Lorentzian region ($a>a_+$) can be estimated
within a WKB formulation. It will depend crucially on the boundary condition
imposed on the wave function of the Universe. Indeed, it is known
that \cite{Vilenkin:1998rp,Bouhmadi-Lopez:2002qz} (see also \cite{Bouhmadi-Lopez:2004mp})
\begin{equation}
\mathcal{A}=\exp(\epsilon 2I), \quad
I=\frac{3\pi}{2\textrm{G}}\int_{a_-}^{a_+}\sqrt{V(a)}\,da,
\end{equation}
where $\epsilon=1$ for the HH wave function and $\epsilon=-1$ 
for the tunnelling (Vilenkin) wave function. Then, the transition amplitude is found as \cite{Vilenkin:1998rp,Bouhmadi-Lopez:2002qz}
\begin{equation}
\mathcal{A}=\exp(\epsilon 2I), \quad
I=\frac{\pi}{2^{\frac32}\textrm{G}}\tilde{K} f,
\label{transition1}
\end{equation}
where
\begin{eqnarray}
f=\frac{\sqrt{1+m}}{\tilde{K}\lambda}\left[\textrm{E}(\alpha_{II})-(1-m)\textrm{K}(\alpha_{II})\right],\,\,
\alpha_{II}=\sqrt{\frac{2m}{1+m}},\nonumber \\ \label{deff}
\end{eqnarray}
with $K(m)$ and $ E(m)$ as complete elliptic integrals of  the first
and second kind, respectively \cite{Gradshteyn}.

\begin{figure}[h]
\begin{center}
\includegraphics[width=0.9\columnwidth]{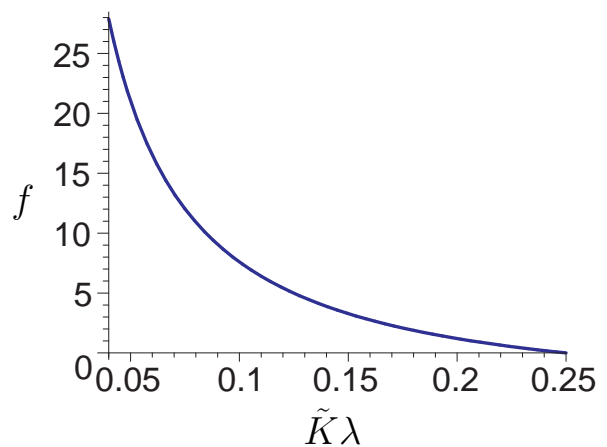}
\end{center}
\caption{Plot of $f$ defined in Eq.~(\ref{deff}) as a function of $\tilde{K}\lambda$, where $f=
\epsilon \sqrt2\textrm{G}/(\pi\tilde{K})\ln\mathcal{A}$. For a
given arbitrary amount of radiation (a given $\tilde{K}\neq 0$), the
preferred value for the cosmological constant by the tunnelling (Vilenkin)
boundary condition (i.e., $\epsilon=-1$) minimises $f$. On the other
hand, the preferred value for the cosmological constant by the
Hartle-Hawking boundary condition, (i.e., $\epsilon=1$) maximises
$f$.}\label{amplitudeplot}
\end{figure}

\begin{figure}[h]
\begin{center}
\includegraphics[width=0.95\columnwidth]{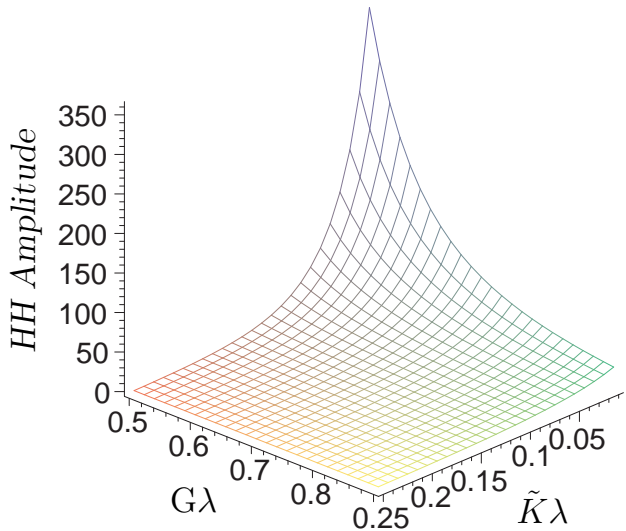}
\end{center}
\caption{Plot of the transition amplitude, $\mathcal{A}$, for the HH boundary condition as a
function of $x=\tilde{K}_a\lambda_a=\tilde{K}\lambda$ and
$\lambda_a$, where $\tilde{K}_a$, $\lambda_a$ are dimensionless
variables defined as $\tilde{K}_a=\tilde{K}/\textrm{G}$ and
$\lambda_a=\textrm{G}\lambda$.}\label{p1}
\end{figure}

\begin{figure}[h]
\begin{center}
\includegraphics[width=0.95\columnwidth]{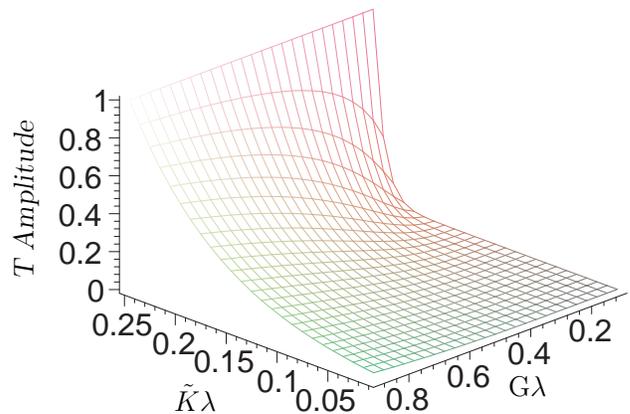}
\end{center}
\caption{Plot of the transition amplitude, $\mathcal{A}$, for the tunnelling (Vilenkin) boundary
condition as a function of $x=\tilde{K}_a\lambda_a=\tilde{K}\lambda$ and
$\lambda_a$, where $\tilde{K}_a$, $\lambda_a$ are dimensionless
variables defined as $\tilde{K}_a=\tilde{K}/\textrm{G}$ and
$\lambda_a=\textrm{G}\lambda$.}\label{p2}
\end{figure}

For a given amount of radiation; i.e. a fixed non-zero value of
$\tilde{K}$, $f$ is a decreasing function of $x\equiv\tilde{K}\lambda$ 
(see Fig.~\ref{amplitudeplot}). Consequently, the HH wave
function suggests a vanishing cosmological constant (see
Figs.~\ref{amplitudeplot} and \ref{p1}); i.e. the
largest amplitude is obtained for $x=\tilde{K}\lambda\rightarrow 0$ or
$\lambda\rightarrow 0$. This is a rather curious aspect. In fact, 
for $\lambda\rightarrow 0$, we 
have $a_+ \rightarrow \infty, ~a_* \rightarrow \infty, ~V(a_*)  \rightarrow \infty$ as
$V(a) \rightarrow a^2 - \tilde{K}$: 
It is an effective harmonic
oscillator behaviour for a zero energy state, as the height of $V$
increases to infinity and a tunnelling endpoint goes to infinity
as well. Hence, ``where'' does the quantum Universe tunnels to in
the case of $\lambda \rightarrow 0$? On the other hand, the 
tunnelling (Vilenkin) wave function
indicates  that 
the largest 
amplitude is obtained for $x\rightarrow 1/4$; i.e. a cosmological 
constant such that  $\tilde{K}\lambda\rightarrow 1/4$, where both
turning points coincide and consequently there is no tunnelling (see
Figs.~\ref{amplitudeplot} and \ref{p2}).
In addition,  the maximum of the potential $V(a)$ also decreases to zero as shown in 
Fig.~\ref{p9}. In essence, there is ``no potential'' to tunnel from
and the evolution will not correspond to a 
quantum mechanical transition \cite{Rubakov:1984bh}.

The above asymptotic features  are based on an arbitrary amount of radiation
as long as  $0<x<1/4$. If one considers instead a
small amount of radiation such that  $x=\tilde{K}\lambda\ll 1$, then
$a_+^2\sim 1/\lambda$, $a_-\sim\tilde{K}$ and $V(a_*)\sim1/(4\lambda)$. 
Therefore, it can be easily seen that
\begin{equation}
f\sim \frac{\sqrt2}{\tilde{K}\lambda}\left[1-\frac34\left(4\ln
2-\ln(\tilde{K}\lambda) +1 \right) \tilde{K}\lambda\right],
\end{equation}
for $\tilde{K}\lambda \ll 1$. Then using Eq.~(\ref{transition1}),
it turns out
\begin{eqnarray}
\mathcal{A}\sim\exp
\left\{\left(\epsilon\frac{\Pi}{G\lambda}\right)
\left[1-\frac34\left(4\ln 2-\ln(\tilde{K}\lambda) +1
\right)\tilde{K}\lambda\right]\right\}.\nonumber \\ \label{amplitude1}
\end{eqnarray}
Consequently, the Vilenkin wave function is in agreement with the
tunnelling of the  Universe and suggests a non
vanishing cosmological constant.  In contrast, the
HH wave function will  still favour a vanishing cosmological
constant:  
This constitutes 
the
standard result 
(cf. for example refs.~\cite{Vilenkin:1998rp,Bouhmadi-Lopez:2002qz,Rubakov:1984bh}) 
for the transition amplitude of a closed FRW
Universe filled with a positive vacuum energy density modulated by some corrections due to the presence of 
a small amount of radiation in the Universe.
Regarding these results, it should be added that, 
(i) in order to retrieve 
a proper tunnelling effect from  the Vilenkin wave
function with radiation,  one needs  an upper ``cut off" in
$\tilde{K}\lambda$ (see Figs.~\ref{amplitudeplot} and \ref{p2}), (ii) in
order to predict a non vanishing cosmological from the HH 
wave function one needs a lower ``cut off" in $\tilde{K}\lambda$ 
(see Figs.~\ref{amplitudeplot} and \ref{p1}). 
The inclusion in this analysis of back-reaction 
effects due to metric fluctuations  and quantum effects due to  
vacuum energy, 
constitute recently addressed issues 
(see, e.g., \cite{Sarangi:2005cs,Barvinsky:2006uh}).

In the next section, we describe how the presence 
of radiation can be employed (cf. \cite{Brustein:2005yn}) 
to formulate (within string theory) an initial quantum state 
for the universe.

\section{Brustein and Alwis Thermal boundary condition}\label{sect3}

 Brustein and de Alwis (BA) have proposed a variant
 boundary
condition for the wave function of the Universe
\cite{Brustein:2005yn}: The Universe emerges
from the string era in a thermal state above the HH
vacuum state (where the HH
vacuum state is the HH wave function with $\tilde{K}=0$).
In addition,  self-consistency of their approach requires  that
the highest energy density on the region of interest ($a_-<a<a_+$)
is such that
\begin{equation} \rho(a_-)=CM_s^4<M_s^4.
\label{nn1}
\end{equation}
Therefore, the parameter $C$ satisfies $C<1$. On the other hand, BA consider that $C$ is
related to the number of degrees of freedom, $n_{dof}$, in thermal
equilibrium through\footnote{In the notation introduced in
\cite{Brustein:2005yn}, $b^4=4\pi^2c^4$.}
\begin{equation}
C=\frac{n_{dof}}{b^4}. \label{nn2}
\end{equation}

As described in Ref.~\cite{Sarangi:2005cs} (cf. Eq. (1.10) therein; 
see also Eqs.~(1.2),~(1.3) and (2.14) in \cite{Firouzjahi:2004mx}; also 
briefly pointed out in BA's paper), $n_{dof}$ constitutes the number of light degrees of freedom 
included in the environment (e.g., for just pure gravity we have $n_{dof}=2$ for the two tensor modes).
More precisely, the framework from which  the setting in  
\cite{Sarangi:2005cs} and consequently that of \cite{Brustein:2005yn}, is 
retrieved is that of a (sub)system (with very limited degrees of freedom; e.g., a time 
dependent scale factor and a 
scalar field), with the remaining degrees of freedom (e.g., fluctuations of those fields) 
constituting the environment (which can be integrated over). Tracing over the fluctuations
produces a decoherence term (cf. Ref.~\cite{Claus}) which will depend on 
a specific cut-off. In string theory, this will depend on the (open and closed) string 
spectrum and has to be calculated for {\em each} vacuum, hence 
depending possibly on compactification and dilaton moduli, among other 
more complex possibilities \cite{Sarangi:2005cs,Firouzjahi:2004mx}.
The  inclusion and full treatment of all these features in 
computation of transition amplitudes can be challenging 
as pointed out in  Ref.~\cite{Firouzjahi:2004mx}.

In attempting to include a parameter that would 
represent a string density of states, this is found to be 
dependent  on the specific model. For example
in \cite{Frey:2005jk}, it is  given the density of state for open strings in
the context of D3 branes in Eq.(D2) and for closed strings in Eq.(D5)
(see also Eqs.~(1)-(2) of Ref.~\cite{Abel:2002rs}). On the other
hand, in Ref.~\cite{Brandenberger:1988aj}
another expression is given for the density of state of strings
compactified in a nine-dimensional box (see for example Eq.(A.1) of
the mentioned paper). 

For simplicity, we follow the indications and procedure 
in \cite{Brustein:2005yn}, which adapts from \cite{Sarangi:2005cs,Firouzjahi:2004mx}.

BA approximate the smaller turning point by
\begin{equation} a_-^2\thicksim \frac{1}{2\lambda}. \label{aproxa-}\end{equation}
We would like to point out that this approximation is valid as
long as $4\tilde{K}\lambda\simeq 1 ^-$  (following the notation
introduced in Section \ref{sect2}); i.e. 
 the thermal effect corresponds to a large amount of
radiation where $\tilde{K}$ is close to $\tilde{K}_{max}$ (see Fig.~\ref{p9}). Moreover, it should be noted that  
$4\tilde{K}\lambda\simeq 1^-$ implies  $m\simeq
0$ leading 
to a potential with a 
very small positive height and $a_- \simeq a_+$ (cf. Eq.~(\ref{eq2})).

Then, combining the last equations, i.e.,
\begin{itemize}
\item Requiring the restriction of $4\tilde{K}\lambda\simeq 1^{-}$ (or equivalently $\tilde{K}\simeq\tilde{K}_{max}$, see Fig.~\ref{p9}) which 
leads to $a_-^2 \sim a_+^2 \simeq 1/2\lambda$,
\item Imposing subsequently the bound (\ref{nn1}),
\item Employing (\ref{nn2}),
\end{itemize}
BA obtain\footnote{In the notation herein used, $\nu$ and $\lambda$
are dimensionfull unlike in \cite{Brustein:2005yn}. 
Nevertheless, the ratio $\nu/\lambda$ in our notation coincide 
with the one introduced in \cite{Brustein:2005yn}. The 
correspondence between our notation and the notation used in 
Ref.~\cite{Brustein:2005yn} is presented in a table  at the 
end of Section~\ref{sect1}. By checking Eq.~(26) of BA paper 
\cite{Brustein:2005yn} it turns out that there is a factor of 
2/3 missing. This does not affect the location of the extremum. 
Then, if we plot this result (bearing in mind that BA use a 
different notation for the complete elliptic integrals \cite{X}), 
it turns out that $g$ reaches it maximum when $\nu/\lambda$ is located 
at 0.12, which coincides with our results. However, in \cite{Brustein:2005yn} 
it is written that the maximum of $\nu/\lambda$ is located at $0.19$; i.e. 
for $\lambda\simeq 5.26\nu$ using the most accurate approximation. It is is also stated in \cite{Brustein:2005yn} that it is also possible to get $\lambda\simeq 8\nu$ using a less accurate approximation based on a triangular integrand.}
\begin{eqnarray}
\rho&=&\frac{3\tilde{K}}{8\pi\textrm{G}}\frac{1}{a^4}, \quad
\tilde{K}\simeq \frac{\nu}{\lambda^2}, \label{newro} \\
\nu&\equiv&\frac23 n_{dof}b^{-4}\pi\textrm{G}M_s^4. \nonumber
\end{eqnarray}
In this manner,  the radiation term ``depends" explicitly on the
cosmological constant $\lambda$.

\begin{figure}[h]
\begin{center}
\includegraphics[width=0.9\columnwidth]{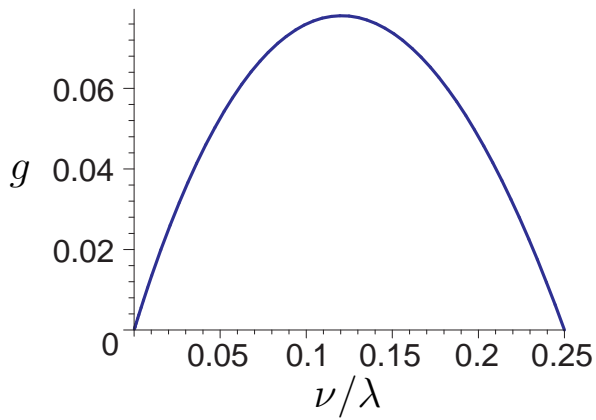}
\end{center}
\caption{Plot of $g$ defined in Eq.~(\ref{defg}) as a function of $\nu/\lambda$, where
$g=\epsilon \frac{\sqrt2}{\pi}\textrm{G}\,\nu \ln\mathcal{A}$. For
a given arbitrary $\nu$, the preferred value for the cosmological
constant by the tunnelling (Vilenkin) boundary condition (i.e. $\epsilon=-1$)
minimises $g$. On the other hand, the preferred value for the
cosmological constant by the HH boundary condition,
(i.e. $\epsilon=1$) maximises $g$.} \label{amplitudeplot2}
\end{figure}

On the one hand, BA  
then conclude that the HH wave function in this case favours a
positive non-vanishing cosmological constant \cite{Brustein:2005yn},
differently to what is exposed in Section \ref{sect2}.
The reason these authors got a different
answer is  because they asked a different question. 
Let us be more precise: 
Instead of looking  for values of $\lambda$ that
maximise the transition amplitude for a fixed value of
$\tilde{K}$, they looked for values of $\lambda$ that maximise the
transition amplitude for a fixed value of $\nu$. Indeed, the
transition amplitude, $\mathcal{A}$,  can be rewritten as
\begin{equation}
\mathcal{A}=\exp(\epsilon 2I), \quad
I=\frac{\pi}{2^{\frac32}\textrm{G}}\frac{1}{\nu}g,
\label{transition2}
\end{equation}
where
\begin{eqnarray}
g=\frac{\nu}{\lambda}{\sqrt{1+m}}\left[\textrm{E}(\alpha_{II})-(1-m)\textrm{K}(\alpha_{II})\right],\,\,
\alpha_{II}=\sqrt{\frac{2m}{1+m}}.\nonumber\\ \label{defg}
\end{eqnarray}
Then, it can be seen that the function $g$ 
is an increasing function of $\nu / \lambda$
until  the ratio $\nu / \lambda$ approaches
$0.12$ and then it starts decreasing (see Fig.~\ref{amplitudeplot2}). 
Consequently, for a fixed value of
$\nu$ the HH wave function favours a non-vanishing cosmological
constant, namely $\lambda\simeq\ 8.33\, \nu$. 

However, let us point out that the result $\lambda\simeq\ 8.33\, \nu$
is not quite  compatible with the approximation 
claimed\footnote{See also footnote 5.} by BA in \cite{Brustein:2005yn}
 (see Eq.~(\ref{aproxa-})); i.e. $\lambda\simeq\ 4\, \nu$. 
Moreover, the use of a WKB approximation is valid as long as 
\begin{equation}
\rm{G}\left|\frac{dV(a)}{da}\right|\ll {\left|V(a)\right|}^{3/2}
\label{WKB1}.
\end{equation}
So, if $4\tilde{K}\lambda\simeq 1^-$, the scale factor in the 
Euclidean region will be of the order of $a_-$ or $a_+$; i.e. 
of the order of $1/\sqrt{2\lambda}$ 
(cf. Eq. (\ref{aproxa-}) and the paragraph below). Consequently, the WKB 
approximation breaks down below the barrier because both 
sides of the  inequality (\ref{WKB1}) vanish. In more detail, the 
WKB approximation can still hold in the Lorentzian regions but 
not in the Euclidean region. Nevertheless this weakness can be 
dealt with. In fact, we will provide in Section \ref{sect4} a broader and 
improved analysis for the thermal boundary condition {\em without}
the restriction $4\tilde{K}\lambda\simeq 1^{-}$. We will also  show how some
 of the conclusions reached in \cite{Brustein:2005yn} can be reversed.

On the  other hand, the tunnelling (Vilenkin) wave function modified by the
thermal effects and for a fixed value of $\nu$, either favours  (i) $\nu /
\lambda \rightarrow 0$ (large $\lambda$ and small $\tilde{K}$)
which is incompatible with the approximation assumed in \cite{Brustein:2005yn} (cf.  
Eq.~(\ref{aproxa-})) or favours (ii) no tunnelling, since   $\lambda\simeq\
4\, \nu$ (or equivalently $4\tilde{K}\lambda\simeq\ 1$). These features are shown in 
Fig.~\ref{amplitudeplot2}.

\section{A Broader Analysis of the Brustein-Alwis  Boundary Condition}\label{sect4}

In this section, we (i) 
widen and improve the application of the thermal boundary condition 
\cite{Brustein:2005yn}  and (ii) 
will see how some of the results obtained in
the previous section are then  modified  when we relax the condition
$4\tilde{K}\lambda \simeq 1^-$ (implicitly assumed in \cite{Brustein:2005yn}). 
We designate the broader setting introduced in this section as {\em generalised
thermal boundary condition} for the wave function. Herein:
\begin{itemize}
\item We will assume instead an arbitrary amount of radiation,
 consistent with a  tunnelling
 of the Universe; i.e.  $\tilde{K}<\tilde{K}_{max}$ (see Fig.~\ref{p9});
\item We  
will not assume the approximation used in Eq.~(\ref{aproxa-}) 
and therefore   we will rather
consider the {\em full} expression (\ref{eq2}) for $a_-$ given in 
Section~\ref{sect2}; 
\item Consequently, we will not restrict ourselves to the 
particular situation corresponding to a large 
amount of radiation;
\item The radiation energy density still reads
\begin{equation}
\rho=\frac{3\tilde{K}}{8\pi\textrm{G}}\frac{1}{a^4},
\label{newro2}\end{equation}
but  now $\tilde{K}$ is related to the cosmological constant
through the more general expression 
\begin{equation}
\tilde{K}= \frac{4\nu\lambda^{-2}}{(1+4\nu\lambda^{-1})^2},
\label{eq20}\end{equation}
while $\nu$ is defined as before; i.e.
\begin{equation}\label{eq21-a}
\nu\equiv\frac23 n_{dof}b^{-4}\pi\textrm{G}M_s^4.
\end{equation}
\end{itemize}
\begin{figure}[h]
\begin{center}
\includegraphics[width=0.9\columnwidth]{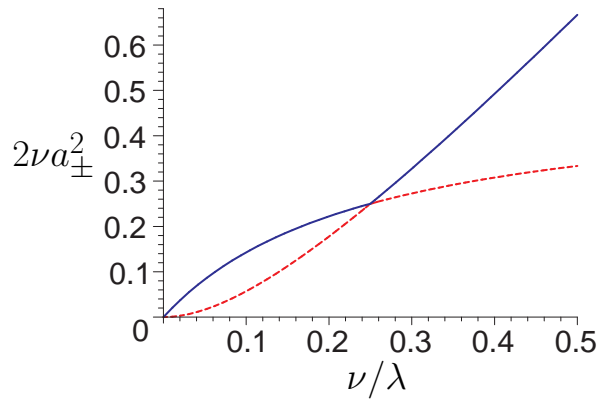}
\label{p5}
\end{center}
\caption{Plot of the dimensionless turning points $2\nu a^2_{\pm}
$ as a function of $\nu/\lambda$. The blue solid line corresponds to
the larger turning point $2\nu a^2_{+}$. The red dash-dot line
corresponds to the smaller turning point $2\nu a^2_{-}$. Although
it seems that both turning points coincide for very small values
of $\nu/\lambda$ it is not the case (see the text).
Consequently, even in this case the Universe can tunnel through the 
barrier $V(a)$. For $\nu/\lambda$ approaching $1/4$ both
turning points coincides and 
consequently 
there is no tunnelling.} \label{p5}
\end{figure}
\begin{figure}[h]
\begin{center}
\includegraphics[width=0.9\columnwidth]{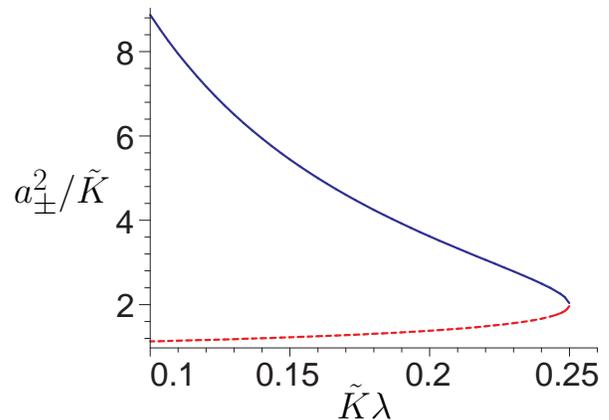}
\end{center} \caption{Plot of the dimensionless turning
points $a^2_{\pm}/\tilde{K}$ as a function of $\tilde{K}\lambda$.
The blue solid line corresponds to the large turning point
$a^2_{+}/\tilde{K}$. The red dash-dot line corresponds to the smaller
turning point $a^2_{-}/\tilde{K}$. For $\tilde{K}\lambda$
approaching $1/4$ both turning points coincide and 
 there is no
tunnelling.} \label{p6}
\end{figure}

\begin{figure}[h]
\begin{center}
\includegraphics[width=0.9\columnwidth]{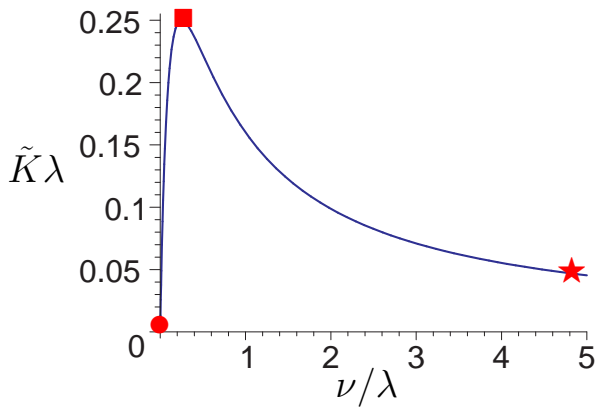}
\end{center}
\caption{Plot of $\tilde{K}\lambda$ as a function of
$\nu/\lambda$. $\nu/\lambda$ is bivalued as a function of
 $\tilde{K}\lambda$ except in the particular case $4\nu/\lambda=1$ or
equivalently $4\tilde{K}\lambda=1$. There is a
 one to one correspondence between the circle, the square and the star 
in this plot and the ones in Fig.~\ref{amplitudeplot4}.} \label{amplitudeplot3}
\end{figure}

\begin{figure}[h]
\begin{center}
\includegraphics[width=0.9\columnwidth]{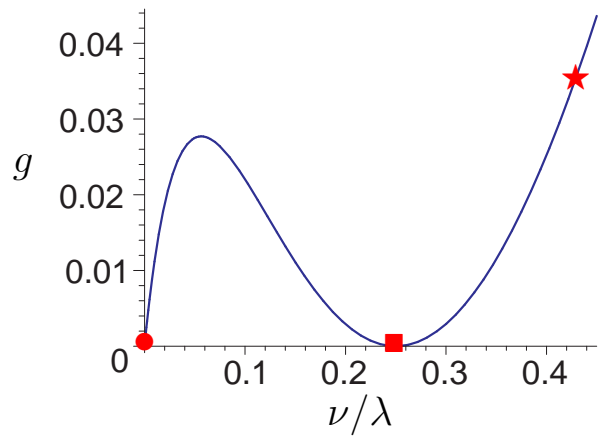}
\end{center}
\caption{Plot of $g$ introduced in Eq.~(\ref{defg}) with the parameter $m$
given in Eq.~(\ref{newm}) as a function of the ratio $\nu/\lambda$. There is a one to one correspondence between the circle, the square and the star in this plot and the ones in Fig.~\ref{amplitudeplot3}. The preferred cosmological situation by the HH wave function is depicted by a star in the plot. The preferred cosmological situations by the tunnelling (Vilenkin)
wave function is depicted by a circle and a square in the figure.}
\label{amplitudeplot4}
\end{figure}

It should  hence be noticed that we recover 
Eq.~(\ref{newro}) using Eqs.~(\ref{newro2})-(\ref{eq21-a}) 
when $4\nu/ \lambda$ is approaching 1.  In this 
limit, both turning points come closer. Indeed, $a_-$ and $a_+$  
coincide at $4\nu/ \lambda=1$ (see Fig.~\ref{p5}). Equivalently, 
in this case $4\tilde{K}\lambda$ is approaching 1 being the case 
where both turning points overlap, inducing no tunnelling. This 
limiting case is represented by a square in Figs.~\ref{amplitudeplot3} 
and \ref{amplitudeplot4} (see also Fig.~\ref{p6}).

From  Fig.~\ref{p5}, it looks like that there is another physical situation
where both turning points would coincide, corresponding to a large cosmological constant
($\nu/\lambda\rightarrow 0$) and a small amount of radiation as measured by the parameter $\tilde{K}$
($\tilde{K}\lambda\rightarrow 0$).  This limiting case is represented by a circle in Figs.~\ref{amplitudeplot3} and \ref{amplitudeplot4}. However, in this case it can be easily seen that although the turning points are small
\begin{equation}
a_-^2\sim \frac{4\nu}{\lambda^2}\sim\tilde{K},\quad a_+^2\sim
\frac{1}{\lambda},
\label{21}\end{equation}
their relative ratio is very large
\begin{equation}
\frac{a_+^2}{a_-^2}\sim\frac{1}{\tilde{K}\lambda}\gg 1.\label{22}
\end{equation}
Consequently, the Universe may still tunnel from the first
Lorentzian region to the larger one. This comment will
be important  for the tunnelling (Vilenkin) proposal (see below and Fig.~\ref{amplitudeplot4}).

Within the broader range employed in this section, the 
relevant feature is that the transition amplitude as a
function of $\nu / \lambda$ will  be unlike the one deduced 
in \cite{Brustein:2005yn}. This is represented 
in  Fig.~\ref{amplitudeplot4}, which is indeed quite
distinct from Fig.~\ref{amplitudeplot2}. 
The reason is that 
 the dependence of the radiation energy density on 
$\lambda$ is modified with respect to the one used in Section~\ref{sect3} 
(cf.~Eqs.~(\ref{newro}) and (\ref{eq20})), although the transition amplitude 
is  given by Eqs.(\ref{transition1})
or  (\ref{defg}). 
Our generalised expressions for the transition amplitudes involve 
the use of Eqs.~(\ref{transition1})-(\ref{deff}) or Eqs.~(\ref{transition2})-(\ref{defg}), but with the parameter $m$  given
by 
\begin{equation}
m=\frac{|1-4\frac{\nu}{\lambda}|}
{1+4\frac{\nu}{\lambda}}\label{newm}.
\end{equation}
and therefore the parameter $\alpha_{II}$ in Eq.~(\ref{deff}) or
Eq.~(\ref{defg}) is modified accordingly. Moreover, 
the modification of the transition amplitude with respect to BA case
is due to the modified Eq.~(\ref{eq20}) which is different from Eq.~(\ref{newro}).
A plot  of $g$ relevant for our new  generalised expressions for the
transition amplitude is given in Fig.~\ref{amplitudeplot4}.

In such context, a pertinent 
question follows:  What will be then the   most likely value of the cosmological
constant for a given value of $\nu$? 

Regarding the HH wave function, it now favours a 
vanishing cosmological constant ($\nu /\lambda  
\rightarrow  \infty$) and $\tilde{K} \sim 1
/(4\nu)$. This physical case is represented 
schematically by a star in Figs.~\ref{amplitudeplot3} 
and \ref{amplitudeplot4}. The turning points can be approximated by 
\begin{equation}
a_-^2\sim \frac{1}{4\nu}\sim\tilde{K},\quad a_+^2\sim
\frac{1}{\lambda},
\label{23}\end{equation}
and, in particular, it turns out that 
\begin{equation}
\frac{a_+^2}{a_-^2}\sim\frac{1}{\tilde{K}\lambda}\gg 1.
\label{24}\end{equation}
Moreover, it can be easily shown that the transition amplitude 
in this case can be approximated by
\begin{eqnarray}
\mathcal{A}&\sim&\exp\left\{\frac{\pi}{\rm{G}{\lambda}}\left[1
-\frac{3}{16}\left(6\ln2-\ln\frac{\lambda}{\nu}+1\right)\frac{\lambda}{\nu}\right]\right\}\nonumber\\
&\sim& \exp\left\{\frac{\pi}{\rm{G}{\lambda}}\left[1
-\frac{3}{4}\left(4\ln2-\ln(\tilde K {\lambda})+1\right){\tilde K}{\lambda}\right]\right\}.\nonumber\\\label{amplitude2}
\end{eqnarray}
In this manner, the role of the HH wave function 
and subsequent transition amplitude is returned to 
its ``original'' implication, with  the thermal  boundary 
condition being  implemented in a  fully consistent 
manner and not restricted to a narrow (perhaps not fully 
valid) 
limit. Notice as well that the
expressions (\ref{21}) [(\ref{22})] and (\ref{23}) [(\ref{24})] 
coincide 
as a function of $\tilde{K}$,  though they correspond to different 
physical situations. Indeed, the limiting case corresponding to 
Eqs.~(\ref{21}) and (\ref{22}) [(\ref{23}) and (\ref{24})] is 
represented by a circle [star] in Figs.~\ref{amplitudeplot3} 
and \ref{amplitudeplot4}. Furthermore, even though the maximum 
height of the potential in both cases is given by the same 
expression $V(a_*)\sim 1/(4\lambda)$, in the HH case $V(a_*)$ is 
very large because $\lambda$ is very small,  while in the other case, 
depicted by a ``circle'' in Figs.~\ref{amplitudeplot3} and \ref{amplitudeplot4}, 
the height of the potential is very small because $\lambda$ is very large. In 
addition, it turns out that in both cases after the tunnelling, the size of the 
Universe does not depend on $\nu$ or on the number of ``degrees of freedom'', 
although the maximum size of the first Lorentzian region  depends on $\nu$. The 
expressions (\ref{21}) [(\ref{22})] and (\ref{23}) [(\ref{24})] coincide 
when taken 
as a function of $\tilde{K}$, because $\nu/\lambda$ is bivalued as 
a function of
$\tilde{K}\lambda$ (except in the particular case $4\nu/\lambda=1$ or
equivalently $4\tilde{K}\lambda=1$) (cf. Fig. \ref{amplitudeplot3}).

Concerning the tunnelling (Vilenkin) wave function, 
it  favours two possible physical situations depicted 
by a circle and a square in Figs.~\ref{amplitudeplot3} 
and \ref{amplitudeplot4}. On the one hand, the ``circle'' option 
corresponds to a large cosmological constant ($\nu / \lambda \rightarrow 0$) and 
a small amount of radiation as measured by $\tilde{K}$ ($\tilde{K}
 \lambda \rightarrow 0$), with the turning points being 
given in Eq.~(\ref{21}) and 
where the transition amplitude can be approximated 
by\footnote{It should be noticed that the 
second expression of the amplitude given in 
Eqs.~(\ref{amplitude2}) and (\ref{amplitude3}) coincide 
with the one given in Eq.~(\ref{amplitude1}). The reason 
is again due to the fact that $\nu/\lambda$ is bivalued as a 
function of $\tilde{K}\lambda$ (except in the particular case 
$4\nu/\lambda=1$ or equivalently $4\tilde{K}\lambda=1$).}
\begin{eqnarray}
\mathcal{A}&\sim&\exp\left\{-\frac{\pi}{\rm{G}{\lambda}}\left[1
-{3}\left(2\ln2-\ln\frac{\nu}{\lambda}+1\right)\frac{\nu}{\lambda}\right]\right\}\nonumber\\
&\sim& \exp\left\{-\frac{\pi}{\rm{G}{\lambda}}\left[1
-\frac{3}{4}\left(4\ln2-\ln(\tilde K {\lambda})+1\right){\tilde K}{\lambda}\right]\right\}.\nonumber\\ \label{amplitude3}
\end{eqnarray}
On the other hand, the ``square'' option implies no tunnelling, that is,  $4 \nu / \lambda
\rightarrow 1$ or equivalently $4\tilde{K} \lambda \rightarrow 1$; i.e. both turning points coincide. In this case, the transition amplitude can be written as 
\begin{eqnarray}
\mathcal{A}&\sim& \exp\left[-\frac{3\pi^2}{32\sqrt 2}\frac{1}{\rm{G}\lambda}\left(1-4\frac{\nu}{\lambda}\right)^2\right]\nonumber\\
&\sim& \exp\left[-\frac{3\pi^2}{32\sqrt 2}\frac{1}{\rm{G}\lambda}\left(1-4\tilde{K} {\lambda}\right)^2\right].
\label{Namp3a}
\end{eqnarray}
It is not possible at this stage to indicate 
which of these two possibilities is more likely. In order to shed some light on
this question, we will employ   DeWitt's argument 
\cite{DeWitt1,Davidson:1999fb,Bouhmadi-Lopez:2004mp} in Section \ref{sect5}.

Before proceeding, let us clarify the following concerning our 
results in this section:

\begin{itemize}
\item Eq.~(\ref{21}) (corresponding to a circle in Figs.~\ref{amplitudeplot3} and \ref{amplitudeplot4}) corresponds to
the turning points for small amount of radiation as measured by
$\tilde{K}$ and a large $\lambda$. The transition amplitude in this
limiting case is given in Eq.~(\ref{amplitude3}). This means we discussed the 
case with  radiation density being less than the amount which saturates the limit 
for semiclassical consistency (cf.~Ref.~\cite{Brustein:2005yn}).

\item The amount of radiation can be made smaller than the above 
 (with respect to $M_s^4$), by having a smaller constant $C$.
But this situation does not change  our results significantly.

\item Eq.~(\ref{23}) (corresponding to a star in Figs.~\ref{amplitudeplot3} and \ref{amplitudeplot4}) corresponds to the turning points for
$\nu/\lambda\rightarrow\infty$; i.e. a small cosmological constant and
$\tilde{K}\sim 1/(4\nu)$. The last condition comes from taking the
limit $\nu/\lambda\rightarrow\infty$ in Eq.~(\ref{eq20}). The transition amplitude in
this limiting case is given in Eq.~(\ref{amplitude2}).

\end{itemize}

When the radiation density exceeds the semiclassical 
saturation bound, specific string effects would have to be considered \cite{Frey:2005jk,Abel:2002rs,Brandenberger:1988aj,Leblanc:1988eq,Lee:1997iz}.                                                         

In the heterotic string theory \cite{Leblanc:1988eq} below the
Hagedorn temperature essentially only the massless
excitations (radiation) contribute to the energy density of the gas of the
heterotic string excitations. As one approaches the Hagedorn temperatures 
i) the radiation density gets larger and larger, meaning larger and larger
values for $C$, which implies a smaller scale factor and therefore a larger
temperature and ii) the
energy density associated with the massive strings is no longer
negligible. Therefore, the
expression for the energy density that fills the universe would be
radiation plus  other components. A subsequent 
analysis will have to include the energy density of the massive states. 

In a complementary context, from the thermodynamics of strings in
the background of D-branes, it can 
be also  assumed that open string dominates over closed strings \cite{Frey:2005jk,Lee:1997iz}. 
For example, particularising to the case of $D3$
 branes,  the
equation of state for the open string gas in the Hagedorn regime was given in 
\cite{Frey:2005jk,Abel:2002rs}. This equation of state is a specific case of
a generalised Chaplygin gas equation of state where 
the strings will redshift like dust; i.e. pressure
less matter. In this case a dust term must be added 
to the potential $V(a)$ (see Eq.~(\ref{WdW})) in addition to the radiation term and
the cosmological constant.
The gas of strings will modify the turning points. 
If $\tilde{K}$ acquire its
maximum allowed value in absence of a gas of strings then there are no
longer turning points and therefore no tunnelling. On the other hand,
if $\tilde{K}$ is small enough then there are two turning points
(closer than in absence of a gas of strings) and the transition
amplitude given by Eq.~(\ref{transition1}) will be smaller. This conclusion is based in comparing a
situation with and without a gas of strings  but with the same amount of radiation, i.e. same $\tilde K$.
However, this situation is not very realistic because the parameter $\tilde K$ should be smaller
in presence of a gas of strings.

Of course,  the transfer of energy from
open strings into radiation (in the reheating phase) can be introduced.
Then, the energy densities of the string gas and
radiation will depend on the Hubble rate (see Ref.~\cite{Frey:2005jk}) which takes into  
account the energy loss of strings into radiation.
Now, because the energy densities depends on the Hubble rate, $H$, the Friedmann
equation is different (it is cubic in $H$). This fact will modify the
Hamiltonian constraint and therefore induce a modification of the
Wheeler-DeWitt equation. This analysis  is out of the scope
of this paper. In any case, we would like to stress that in the present paper, we are interested in a  situation below the Hagedorn temperature like in Ref.~\cite{Brustein:2005yn}.

\section{DeWitt's argument and the Brustein and Alwis boundary condition}\label{sect5}

A pertinent feature concerning the minisuperspace of a 
closed FRW Universe with 
radiation and a positive cosmological constant is that 
a classically allowed (Lorentzian) region may exist for $0<a<a_-$
(cf. Fig.~\ref{p9}  and Eq.~(\ref{eq2})). This is of 
relevance due to the following. 
It can be shown that for small values of the scale factor the
corresponding 
Ricci curvature is not well defined. Indeed, it
diverges\footnote{However, the Ricci scalar, $R$, is well defined, for a
radiation filled Universe with a positive cosmological constant.
  $R$ is constant and positive.}. 
In order to deal quantum mechanically with this situation 
an interesting procedure was advocated 
by B. DeWitt \cite{DeWitt1}, which has been recently 
used in Refs.~\cite{Davidson:1999fb,Bouhmadi-Lopez:2004mp}. 
In more detail, DeWitt's argument  involves the assumption  that the 
wave function vanishes at the singularity; i.e. at $a=0$ for our FRW model.

The DeWitt's argument applied to BA wave function yields 
\cite{Bouhmadi-Lopez:2004mp}

\begin{equation}
\frac{3\pi}{2\textrm{G}}\int_0^{a_-}\,\sqrt{-V(a)}\,da-\frac{\pi}{4}=n\pi,\quad
n\in \mathbb{N}. \label{DWcondition}\end{equation}
The previous condition results in (see for example
\cite{Bouhmadi-Lopez:2002qz})
\begin{equation}
\frac{\pi}{2^{\frac32}\textrm{G}\lambda}\sqrt{1+m}
\left[\textrm{E}(\alpha_{I})-m\textrm{K}(\alpha_{I})\right]-\frac{\pi}{4}=n\pi,\label{DWcondition2}
\end{equation}
where
\begin{equation}
\alpha_{I}=\sqrt{\frac{1-m}{1+m}}.
\label{alphaI}\end{equation}
We recall that the thermal boundary condition under the description 
of Ref.~\cite{Brustein:2005yn} (cf. Section 3)     suggests 
$\lambda\thicksim\ 8.33\, \nu$. Consequently, we obtain\footnote{See Refs.~\cite{Melnikov,Claus}, for  previous proposals for quantisation conditions of the cosmological constant. We thank C.~Kiefer for pointing out to us these references.}

\begin{equation}
\textrm{G}\lambda \thicksim 0.4\sqrt{2} (1+4n)^{-1},
\label{quanta1}\end{equation}
which implies not only that the weaker is gravity (smaller $G$),
the larger is the cosmological constant (for a fixed value of $n$)
 but also that there is a quantification relation between $G$ and
 $\lambda$.

As we already mentioned in Section \ref{sect3}, $\lambda\thicksim\ 8.33\, \nu$ is not
fully in agreement with the approximation made in 
Eq.~(\ref{aproxa-}): $\lambda\thicksim\ 4\, \nu$, i.e. 
with a large amount of radiation (see Fig.~\ref{p9}) 
implicit on the BA thermal boundary condition  context (see the third line
before Eq.~(13) in \cite{Brustein:2005yn}). If we take into account
the approximation in Eq.~(\ref{aproxa-}) or equivalently $\lambda\thicksim\ 4\, \nu$,
then it turns out that
\begin{equation}
\textrm{G} \lambda \thicksim \sqrt{2}(1+4n)^{-1}. \label{quanta2}
\end{equation}
Consequently, we obtain a fairly similar conclusion.

Another  interesting  result follows for the cosmological constant 
and the string mass scale. In fact, Eq.~(\ref{newro}) and  the 
quantification relation (\ref{quanta2})
(the final result between $M_s$ and $\lambda$ is slightly modified
for Eq.~(\ref{quanta1})) implies
\begin{equation}
\frac{2^{3/2}}{3}\frac{n_{dof}}{b^{4}}\pi M_s^4 
\thicksim  \lambda^2 \left(n+\frac14\right).\label{quanta3}
\end{equation}
The larger is $M_s^2$, the larger is $\lambda$ (for  fixed values of $n$, $b$ and $n_{dof}$). In addition, the larger is $n$, the smaller is the cosmological constant (given $M_s^2$, $b$ and $n_{dof}$ fixed).

Finally, we would like to point out that the previous
relations given in Eqs.~(\ref{quanta2}) and
(\ref{quanta3}) implies a relation between $M_s$, $M_p$
($\textrm{G}=M_p^{-2}$) and $n_{dof}$:
\begin{equation}
M_{p}^4\thicksim \frac{2^{5/2}}{3}\frac{n_{dof}}{b^4}\pi M_s^4
(1+4n). \label{quanta4}\end{equation} 
It turns out that the larger is $M_s$, the smaller is the number of degrees of freedom in thermal equilibrium given by $n_{dof}$ (for  fixed values of $n$, $b$ and $M_p$). This result is in agreement with Eqs.~(\ref{quanta2}) and (\ref{quanta3}) under the same assumptions; i.e. for  fixed values of $n$,  $b$ and $M_p$.

Let us now apply DeWitt's argument \cite{DeWitt1} (see also  \cite{Davidson:1999fb,Bouhmadi-Lopez:2004mp}) 
to establish which of the two possibilities favoured by the tunnelling wave functions 
(see  Eqs.~(\ref{amplitude3})and (\ref{Namp3a})) is more probable once the generalised thermal boundary condition is imposed (cf. Section~\ref{sect4}).

It is then crucial to remark the following. 
The modified tunnelling (Vilenkin) wave function by the 
thermal effect proposed in \cite{Brustein:2005yn} (and assuming  
the approximation used in Eq.(\ref{aproxa-}) which is  implicit in  
\cite{Brustein:2005yn}) is incompatible with DeWitt's argument. The 
reason is that such  wave function favours no tunnelling; i.e. the 
turning points are of the same order
 of magnitude ($a_{\pm}^2 \thicksim 1/(2\lambda)$),  while DeWitt's argument
applied to the tunnelling wave function requires a ``wide barrier'', more precisely
\cite{Bouhmadi-Lopez:2004mp}
\begin{equation}
1\ll
\exp\left[\frac{3\pi}{2\textrm{G}}\int_{a_-}^{a_+}\sqrt{V(a)}\;da\right].
\label{tunnelingdewitt}\end{equation} 
Consequently, if $a_{\pm}^2
\thicksim 1/(2\lambda)$, the previous condition cannot hold and we
cannot apply the DeWitt argument to the tunnelling wave function
within the context of \cite{Brustein:2005yn} (i.e., of Section 3), 
which reads \cite{Bouhmadi-Lopez:2004mp}
\begin{equation}
\frac{3\pi}{2\textrm{G}}\int_0^{a_-}\sqrt{-V(a)}\;da=
\frac{3\pi}{4} + \tilde{n}\pi,\quad \tilde{n}\in\mathbb{N};
\label{tunnellingquanta}
\end{equation}
or equivalently
\begin{equation}
\frac{\pi}{2^{\frac32} \textrm{G}\lambda}\sqrt{1+m}\left[\textrm{E}(\alpha_I)-m\textrm{K}(\alpha_I)\right]-\frac{3\pi}{4}=\tilde{n}\pi,
\end{equation}
where $\alpha_I$ is defined in Eq.~(\ref{alphaI}).

However, it turns out that we can indeed apply 
 the DeWitt's argument to the tunnelling (Vilenkin) wave function 
but within the  ``generalised'' analysis for the thermal 
boundary condition (cf. Section~\ref{sect4}).  
The condition (\ref{tunnelingdewitt})  is required to hold. So firstly, we check 
 if this condition can be fulfilled.

Following the analysis of  Section~\ref{sect4}, it turns out that
the tunnelling wave function favours either (i) a large cosmological
constant ($\nu / \lambda \rightarrow 0$) and a small amount of
 radiation as measured by $\tilde K$ ($\tilde{K} \lambda \rightarrow 0$)  
or (ii) that there is no tunnelling, that is, $4 \nu / \lambda
\rightarrow 1$ which implies $4\tilde{K} \lambda \rightarrow 1$. 
These features are depicted in Figs.~\ref{amplitudeplot3} and \ref{amplitudeplot4}.

Condition (ii) is incompatible with the condition (\ref{tunnelingdewitt})
because for (ii) there is no tunnelling. 

Condition (i) is compatible (and therefore chosen) 
within the DeWitt's requirement 
despite that  $\nu / \lambda \rightarrow 0$ implies $g\rightarrow 0$.
For $\nu / \lambda \rightarrow 0$ the function $g$ (related to the transition amplitude (defined in Eq.~(\ref{defg}) where $m$ is given by Eq.~(\ref{newm})) can be approximated by
\begin{equation}
g\thickapprox \sqrt{2}\;\frac{\nu}{\lambda}.
\end{equation}
Consequently, the condition (\ref{tunnelingdewitt}) reads
\begin{equation}
1\ll \exp\left(\frac{\pi M_p^2}{2\lambda}\right),
\end{equation}
which implies that the DeWitt argument is satisfied  as long as the cosmological constant $\lambda$ is much smaller than $M_p^2$. In this case, the quantisation relation (\ref{tunnellingquanta}) reads
\begin{equation}
\frac{3\pi}{2}\frac{\nu}{\textrm{G}\lambda^2}\thicksim\frac{3}{4}+\tilde{n}.
\end{equation}
Substituting  the definition of $\nu$ in the previous equation results on
\begin{equation}
\pi^2\frac{n_{dof}}{b^4} M_s^4\thicksim\left(\frac34+\tilde{n}\right)\lambda^2.
\end{equation}
Consequently, the larger is $M_s^2$, the larger is $\lambda$ (for  fixed values of $\tilde n$, $b$ and $n_{dof}$). In addition, the larger is $\tilde n$, the smaller is the cosmological constant (given $M_s^2$, $b$ and $n_{dof}$ fixed).

In order to get a relation between $M_p$, $M_s$ and $n_{dof}$, it
is necessary to go to the next order in the expansion of 
Eq.~(\ref{tunnellingquanta}) as a function $\nu/\lambda$.

We recall that the DeWitt's argument is
applied in the Lorentzian regions $0<a<a_-$ where the tunnelling
(Vilenkin) wave function is a superposition of ingoing and outgoing
modes. In fact, the tunnelling wave function for a closed FRW Universe filled with
radiation and a positive cosmological constant in the Lorentzian
region $0<a<a_-$ is a linear combination of ingoing and outgoing modes
and {\em not} a sum of decaying and growing modes. To be more precise,
 from a
classical point of view the region $0<a<a_-$ describes a Universe that
starts at $a=0$ and expands up to a maximum radius $a_-$,  as depicted
in Fig.~1 of Ref.~\cite{Vilenkin:1998rp}. Therefore, the tunnelling wave function
on this region cannot be formed from decaying and growing modes, in
particular close to $a=0$, because this
region does not correspond to a Euclidean region. It is in absence of
radiation (which is not the physical situation analysed in this paper)
that the tunnelling wave function corresponds to a sum of growing and
decaying modes for $0<a<1/(2\lambda)$, in particular close to $a=0$,
as is illustrated in Fig.~2 of \cite{Vilenkin:1998rp}. We refer the reader to Refs.~\cite{Davidson:1999fb,Bouhmadi-Lopez:2004mp} for a detailed
discussion on the DeWitt's argument, the tunnelling wave function and
the Hartle-Hawking wave function.

For completeness, let us see what happens to the set of
parameters once the DeWitt's argument is imposed within HH case once the generalised thermal boundary condition is assumed. In this case
$\lambda/\nu\rightarrow 0$. Using\footnote{Notice that Eqs.~(\ref{DWcondition2}) and (\ref{alphaI}) depend on the amount of radiation only through the parameter $m$. Consequently, these equations can be applied for a large amount of radiation (as is the case in Section~\ref{sect3}) or for an arbitrary amount of radiation (consistent with the tunnelling of the Universe as in Section~\ref{sect4}) by choosing an appropriate expression for the parameter $m$.} Eq.~(\ref{DWcondition2}), where
 now $m$ is given by Eq.~(\ref{newm}) and in the
limiting case $\lambda/\nu\rightarrow 0$, it turns out that
\begin{equation}
M_{p}^2\sim\frac{8}{3\pi}(1+4n)\nu.
\end{equation}
Furthermore, substituting  the definition of $\nu$ in the previous
equation results on
\begin{equation}
M_p^4\sim\frac{16}{9}(1+4n)\frac{n_{dof}}{b^4}M_s^4.
\end{equation}
Notice that the previous expression is fairly similar to
Eq.~(\ref{quanta4}). On the other hand, notice also that the last two
expressions do not depend on the cosmological constant.

\section{Including Moduli Fields}\label{sect6}

Up to now, we have been using a simplified model by considering $\lambda$ constant. 
However, rather than a cosmological constant we expect to have a moduli 
dependent potential $\lambda(\phi)$. Moreover, the ratio $M_s^2/M_p^2$ will
 also depend on $\phi$ (c.f. \cite{Brustein:2005yn} and references therein). 
In the heterotic string theory $M_s^2/M_p^2 = \alpha_{YM}/8$, where $\alpha_{YM}$ is
 the gauge field coupling strength at the string scale. On the other hand, in type
 I string theory $M_s^2/M_p^2 = g\; \alpha_{YM}/4$, where $g$ is the string coupling. 
Our aim in this section is to constrain (at least qualitatively) 
how  the 
Universe 
(in the presence of moduli fields) will behave 
after its tunnelling. 
This will be presented 
within the broader discussion of the 
generalised thermal boundary condition 
introduced in Section~\ref{sect4}.  To be more 
precise, this will allow us to discuss in a more self-consistent manner the issue 
of transition amplitude and corresponding probability distribution, when the 
 presence 
of moduli fields and their potentials is of relevance.

\subsection{``Flat'' moduli potential}

Let us consider a rather particular situation, namely that 
 the moduli potential, $\bar{V}(\phi)\equiv\sigma^2\lambda(\phi)$, is flat enough; i.e. 
$$\left|\frac{d\lambda}{d\phi}\right|\ll \left|\lambda(\phi)\right|,$$
at least just after the tunnelling of the Universe. In other words, 
we take the case (not likely in string theory) that the moduli potential is not 
steep (cf. also \cite{Brustein:2005yn}). Consequently, applying the framework introduced by Vilenkin in \cite{Vilenkin:1987kf}, we can substitute $\lambda$ by $\lambda(\phi)$ in the expressions obtained previously. 

Then adapting the expressions in section~\ref{sect4} consistently, we found for the tunnelling (Vilenkin) wave function that there are two possibilities:

\begin{enumerate}

\item If $\lambda\sim 4\nu$ then
\begin{equation}
\frac{3\pi}{2}\sigma^2{\lambda(\phi)}=\frac{8\pi}{3}\frac{n_{dof}}{b^4}\left(\frac{M_s}{M_p}\right)^4(\phi),
\label{29}\end{equation}
where $\sigma^2={2\rm{G}}/(3\pi)$. By considering that $\lambda$ is 
much smaller than the Planck scale; i.e. $\lambda/M_p^2\ll 1$ where 
${\rm{G}}=M_p^{-2}$ and  as we are working in a semiclassical framework, 
we can conclude that the tunnelling (Vilenkin) wave function 
(once the generalised thermal boundary condition is applied) 
seems to favour  the emergence of a weakly coupled Universe
(at least in the heterotic and type I theories). 

\item If $\nu/\lambda\ll 1$ then
\begin{equation}
%
\frac{2\pi}{3}\frac{n_{dof}}{b^4}\left(\frac{M_s}{M_p}\right)^4(\phi)\ll
\frac{3\pi}{2}\sigma^2{\lambda(\phi)}.
\label{30}\end{equation}
Following the argument of the previous item, 
we can conclude that in this case the tunnelling (Vilenkin) 
wave function (once the generalised thermal boundary condition 
is  applied) seems to favour  the tunnelling of the Universe 
towards a weakly coupled regime.

\end{enumerate}

Finally, as we have shown in Section~\ref{sect4}, the HH wave function (once the generalised thermal boundary condition is  imposed) implies $\nu/\lambda\gg 1$ and consequently:
\begin{equation}
\frac{2\pi}{3}\frac{n_{dof}}{b^4}\left(\frac{M_s}{M_p}\right)^4(\phi)\gg
\frac{3\pi}{2}\sigma^2{\lambda(\phi)}.
\label{31}\end{equation}
Unlike for the tunnelling (Vilenkin) wave function, we cannot in principle conclude that the Universe will tunnel to a weakly coupled regime. Essentially, it is not enough to impose that $\lambda$ is much smaller than the Planck scale to ``predict'' the emergence of a weakly coupled Universe.

\subsection{Moduli fields and the tunnelling probability}

In the previous subsection, we have supposed that (i) 
the moduli potential is flat enough and (ii) the potential 
dominates the total energy density in such a way that the 
tunnelling towards an inflating Universe takes place. However, 
it turns out that the moduli potentials are in general quite steep, 
leading to a dominance of the kinetic energy density of the scalar 
field over its potential energy \cite{Brustein:2004jp}. This condition 
is  incompatible with the emergence of an accelerating Universe. Possible 
exceptions correspond to the only flat regions of the moduli potentials 
(that support a tunnelling of the Universe towards an accelerated regime 
and are characterised by $\bar{V}'/\bar{V}\ll 1$ with $'\equiv d/d\phi^i$ 
and $i$ labelling the different moduli), which constitute limited domains 
close to a positive extremum of the potentials\footnote{\label{footnote7}Of course, 
we are considering that the potential dominates over the 
total energy density after the tunnelling. In particular, 
$\bar{V}$ dominates over the radiation energy density and 
the kinetic energy of the scalar field.} \cite{Brustein:2005yn}. 
Let us consider (as in \cite{Brustein:2005yn}) a moduli potential 
$\bar{V}$ cast in the form of an $N=1$ SUGRA potential, where the 
K\"{a}hler potential involves the real part of the dilaton axion field 
$S_r$ and the real part of the volume modulus $T_r$. These fields 
determine the ratio $({M_s}/{M_p})^4$ as $1/(S_r T_r^3)$ \cite{Brustein:2005yn}.

The moduli $S_r$ and $T_r$ are constrained to satisfy the ``flat enough'' condition near an extremum
\begin{equation}
\bar{V}'/\bar{V}\sim 0.
\label{extrema}\end{equation}
This condition is consistent with an accelerated expansion phase of the Universe after the tunnelling\footnote{See footnote \ref{footnote7}.}. Moreover, the maximisation of the transition amplitude of  the tunnelling (Vilenkin) wave function (once the generalised thermal boundary condition is imposed) implies two possibilities, adapting the reasoning employed for Eqs.~(\ref{29}) and (\ref{30}), respectively:

\begin{enumerate}

\item The moduli potential is then constrained as follows\footnote{The moduli potential $\bar{V}$ is related to the potential $\lambda(\phi)$ introduced in the previous subsection by $\bar{V}=\sigma^2\lambda$.}
\begin{equation}
\bar{V}(S_r,T_r)\sim \frac{16}{9}\frac{n_{dof}}{b^4}\frac{1}{S_r T_r^3}.
\end{equation}
We would like to point out that in this particular case the extremisation condition (\ref{extrema}) comes from imposing the emergence of an accelerating Universe and not from the tunnelling of the Universe towards an inflating Universe. We recall that in this particular case the tunnelling (Vilenkin) wave function predicts ``no tunnelling''. From now on we will disregard this case.
\item The other possibility is that the moduli potential is constrained by
\begin{equation}
\frac{4}{9}\frac{n_{dof}}{b^4}\frac{1}{S_r T_r^3}\ll \bar{V}(S_r,T_r).
\label{modulitunneling}\end{equation}
\end{enumerate}
However, the maximisation of the transition amplitude of the HH wave function (once the generalised thermal boundary condition is imposed) implies (cf. Eq.~(\ref{31}))
\begin{equation}
\bar{V}(S_r,T_r)\ll\frac{4}{9}\frac{n_{dof}}{b^4}\frac{1}{S_r T_r^3}.
\label{moduliHH}\end{equation}
In summary, the generalised thermal boundary condition of Section~\ref{sect4}, 
when applied to the wave function of the Universe,  produces  
{\em less} restrictive conditions than the ones extracted by BA in \cite{Brustein:2005yn}.

\subsection{Application to a KKLT-like model}

Before concluding this section, let us apply the above discussion to a KKLT model \cite{Kachru:2003aw}, where the effective moduli potential is given by \cite{Kachru:2003aw}
\begin{equation}
\bar{V}=\frac{aCe^{-aT_r}}{2T_r^2}\left[W_0+\left(\frac13 T_r a+1\right)Ce^{-aT_r}\right]+\frac{d}{T_r^3}.
\label{kkltpotential}\end{equation}
In this expression $a$, $C$, $W_0$ and $d$ are constant and the rest of moduli have been stabilised (in particular $S_r\sim O(1)$ \cite{Brustein:2005yn}). This potential has a positive maximum separating a positive minimum from the asymptotically vanishing potential corresponding to the tail of the potential where the term $d/T_r^3$ dominates \cite{Kachru:2003aw}.

Let us see if the conditions (\ref{modulitunneling}) and (\ref{moduliHH}) 
can be fulfilled at the tail of the potential. In this case these conditions 
read $\frac{4}{9}\frac{n_{dof}}{b^4}\ll d$ for the tunnelling (Vilenkin) wave function 
and $d\ll\frac{4}{9}\frac{n_{dof}}{b^4}$ for the HH wave function (once the 
generalised thermal boundary condition is imposed). It turns out that 
because\footnote{We are considering the choice of parameters of the  KKLT 
model in \cite{Kachru:2003aw}. \label{footnote8}} 
$d\sim O(10^{-9})$ and $\frac{4}{9}\frac{n_{dof}}{b^4}\sim O(10^{-1})$ 
the tunnelling (Vilenkin) wave function cannot satisfy the required inequality 
while the HH does it\footnote{These results can be understood 
by taking into account that the tunnelling (Vilenkin) wave function prefers a large cosmological constant, while the HH wave function prefers a small cosmological constant (See Section~\ref{sect4}).}. In any case, despite the previous conclusions, the Universe cannot tunnel towards the tails of the potential because the dominant term $d/T_r^3$ is too steep; i.e. the extremisation condition (\ref{extrema}) cannot hold. This result generalises the one obtained by BA in \cite{Brustein:2005yn}.

Finally, let us see if the tunnelling endpoint can correspond to the extremum of the KKLT potential where the extremisation condition (\ref{extrema}) obviously holds. Let us see if the inequalities (\ref{modulitunneling}) and (\ref{moduliHH}) hold in this case. The tunnelling (Vilenkin) wave function prefers values such that 
\begin{eqnarray}
\frac{4}{9}\frac{n_{dof}}{b^4}\ll d + \frac{aCT_re^{-aT_r}}{2}\left[W_0+\left(\frac13 T_r a+1\right)Ce^{-aT_r}\right],\nonumber\\
\end{eqnarray}
while the HH wave function prefers values satisfying the opposite condition. It turns out that the right hand side of the previous inequality at the maximum of the potential is of order\footnote{see footnote \ref{footnote8}.} $O(10^{-9})$, while the left hand side is of  order $O(10^{-1})$. Consequently, the tunnelling (Vilenkin) wave function does not satisfy the previous inequality at the maximum of the potential and even less at the minimum of the potential; i.e. in this case the tunnelling endpoint cannot correspond to an extremum of the potential. Again, this result is not surprising because the tunnelling (Vilenkin) wave function prefers a large energy density while the extremum of the potential (\ref{kkltpotential}) are very small. Applying a similar argument to HH wave function (once the generalised thermal boundary condition is imposed), we can conclude that the tunnelling endpoint can correspond to an extremum of KKLT potential (\ref{kkltpotential}).

\section{Discussion and Conclusions}\label{sect7}

The framework of quantum cosmology 
\cite{Vilenkin:1983xq,Hartle:1983ai,Linde:1983cm} 
allows to establish a probability distribution for 
the (dynamical) parameters that characterise the 
(multi)-universe associated with the landscape of solutions in string theory \cite{Bousso:2000xa,Douglas:2003um,Susskind:2003kw,Polchinski2:1998rr,Polchinski:2006gy}. Of course, this probability 
distribution depends crucially on the boundary 
condition imposed on the wave function of the Universe.

In this context, Brustein and de Alwis (BA) have recently advanced 
a thermal boundary condition for the wave function of the Universe 
\cite{Brustein:2005yn}. It is proposed that the decay of all the string excited 
states created a primordial thermal gas of radiation. 
Subsequently,  the Universe emerged from the string era 
in a thermal state above the HH vacuum, with an assigned 
probability of tunnelling into  the landscape 
\cite{Bousso:2000xa,Douglas:2003um,Susskind:2003kw}. 
The essential new ingredient in the setting of \cite{Brustein:2005yn} is 
that the cosmological constant ``depends'' on the amount of radiation 
(cf. Eq.~(\ref{newro})). As a consequence of this fact, the thermal 
boundary condition proposed in  \cite{Brustein:2005yn} switches the 
role of the HH and tunnelling (Vilenkin) wave functions. More precisely, in the sense 
that the HH wave function 
(once the thermal boundary condition of \cite{Brustein:2005yn} is imposed) prefers a 
non-vanishing cosmological constant larger than the one 
preferred by the tunnelling (Vilenkin) wave function.

One of our main contributions in this paper (cf. Sections~\ref{sect3} 
and \ref{sect4}) has been to  critically analyse the transition 
amplitude, as discussed in \cite{Brustein:2005yn}, for a closed 
radiation-filled FRW minisuperspace with a positive cosmological 
constant, either for the HH \cite{Hartle:1983ai} 
or the tunnelling (Vilenkin) boundary conditions
 \cite{Vilenkin:1983xq,Vilenkin:1998rp}. 
As a consequence, we have shown that the framework introduced  
in  \cite{Brustein:2005yn} correspond to a very narrow application 
of the proposed thermal boundary condition.

In more detail,  we have proven that the thermal 
boundary condition used in \cite{Brustein:2005yn} 
corresponds to the particular case or specific choice where the amount 
of radiation present in the Universe 
is very large\footnote{See Fig.~\ref{p9}.}; i.e. 
\begin{equation}
\tilde{K}\lambda\sim\frac{\nu}{\lambda}\sim \frac14,\label{conclusion1}
\end{equation}
where $\tilde{K}$ and $\nu$ are parameters 
that measure the amount of radiation (see  Eq.~(\ref{newro})) 
and $\lambda$ is a rescaled cosmological constant, although 
maintaining the possibility for the Universe to tunnel. In addition, 
some of the predictions for the HH wave function within the framework 
of \cite{Brustein:2005yn} are not totally compatible with the approximation 
used therein (cf. Section~\ref{sect3}). Furthermore, the WKB approximation 
is applied in a  not fully cohesive way.

In 
Section~\ref{sect4} 
we proposed instead a generalised thermal boundary condition, 
namely  
 by allowing an arbitrary amount of radiation consistent
 with the tunnelling of the Universe; i.e. \mbox{$0<4\tilde{K}\lambda<1$.} 
Consequently, the relation (\ref{conclusion1}) is replaced by the consistent 
and more general Eq.~(\ref{eq20}). We have then concluded that the preferred 
value of the cosmological constant induced by the HH wave function (once the 
thermal boundary condition is applied properly) is a vanishing 
cosmological constant. In addition, we have shown that the 
preferred value of the cosmological constant by the tunnelling 
(Vilenkin) wave function (once the generalised thermal boundary condition 
is applied, cf. Section \ref{sect4}) is a non-vanishing cosmological constant, 
which can be (i) very large or (ii) proportional to the amount of 
radiation present in the Universe as measured by the parameter 
$\nu$ (see Eqs.~(\ref{newro2})-(\ref{eq21-a})). In order to select 
one of these two possibilities for the tunnelling (Vilenkin) wave function, we 
employed the DeWitt's argument \cite{DeWitt1,Davidson:1999fb,Bouhmadi-Lopez:2004mp} 
(cf. Section~\ref{sect5}), since  
there is a curvature singularity at small scale factors. 
It turns out that the preferred value of the cosmological 
constant in this case is a large one. Moreover, this condition 
implies a small amount of radiation (as measured by the parameter 
$\tilde K$) allowing consequently the tunnelling of the Universe.

In addition,   DeWitt's argument \cite{DeWitt1,Davidson:1999fb,Bouhmadi-Lopez:2004mp} 
have also enabled us to  find that the string and gravitational  parameters  become related through an expression involving an integer $n$, suggesting a quantisation relation for some of the involved parameters.

We have also applied the generalised thermal boundary condition 
to more realistic models, where rather than a cosmological constant 
we have a moduli dependent potential (cf. Sections~\ref{sect6}). We have concluded 
that the tunnelling (Vilenkin) wave function seems to favour the emergence 
of a weakly coupled Universe. In contrast, for the HH wave function we cannot 
in principle conclude that the Universe will tunnel to a weakly coupled regime. 
Moreover, we have also supplied conditions for the moduli, 
with the assistance of a  potential extracted from  N=1 SUGRA, such that 
a tunnelling towards an inflating phase takes place. We have obtained  
that the generalised thermal boundary condition, when applied to the 
wave function of such a Universe, produces less restrictive conditions 
than the one proposed by BA in \cite{Brustein:2005yn}. In particular, 
for a KKLT model \cite{Kachru:2003aw},  the Universe cannot tunnels towards 
the tail of the potential because of its steepness. Moreover, the steepness
 of the potential would not support an accelerating Universe. This result 
apply for the HH and tunnelling (Vilenkin) wave functions, once the 
generalised thermal boundary condition is applied,  generalising the 
one obtained in \cite{Brustein:2005yn}. In addition, we have also shown 
that the tunnelling end points cannot correspond to the extremum of the 
potential if we choose the tunnelling (Vilenkin) wave function, supplied 
with the generalised thermal boundary condition. However, the tunnelling 
end points can correspond to the extremum of the potential if we choose 
the HH  wave function supplied with the generalised thermal boundary condition.
It remains to be analysed how these conclusions are modified by considering more 
accurate potentials for KKLT model like the recently advanced in~\cite{Achucarro:2006zf,Parameswaran:2006jh}.

Before concluding, we would like to stress again that one 
of the key element proposed  by BA in \cite{Brustein:2005yn} 
(although applying a restrictive thermal boundary condition)  
is that the radiation fluid parameter $\tilde{K}$ ``depends'' 
on a specific way on the rescaled cosmological constant $\lambda$ 
(see  Eq.~(\ref{newro})). This  lead to a preferred non-vanishing
 value of $\lambda$ when using HH wave function. This result is 
similar to the one obtained by Sarangi and Tye \cite{Sarangi:2005cs} 
(see also \cite{Firouzjahi:2004mx,Sarangi:2006eb}). However, in this 
latter case the dependence of $\tilde{K}$ on $\lambda$ has a completely
 different physical origin. In fact, it comes from the inclusion of 
decoherence effects due (for example) to metric fluctuations 
\cite{Sarangi:2005cs}. It turns out that the quantum 
fluctuations during the spontaneous creation of the Universe 
generate some radiation,  which depends on $\lambda$ 
(after integrating out the perturbative modes 
appropriately and allowing for  back-reaction). 
More recently, the back-reaction effects on the 
cosmological landscape scenario have also been 
analysed by Barvinsky and Kamenshchik  \cite{Barvinsky:2006uh}. 
Their results again lead to a lower bound on the allowed values 
of the cosmological constant \cite{Barvinsky:2006uh}. More 
importantly, they suggest a mechanism that eliminates the infrared 
catastrophe of a small cosmological constant in Euclidean quantum cosmology
 (see also \cite{Sarangi:2005cs}). 
Consequently, it seems very important to include these back-reaction effects in particular when we apply the generalised thermal boundary condition. We leave this interesting issue for a future work.

\section*{Acknowledgments}

We thank L.~J.~Garay and D.~Wands for 
useful discussions. The authors are also grateful to  P.~F.~Gonz\'{a}lez-D\'{\i}az, A.~Kamenshchik, C.~Kiefer  and J.~Ward
for their helpful comments and R.~Brustein and A.~de~Alwis for correspondence. MBL acknowledges the support 
of CENTRA-IST BPD (Portugal) as well as the FCT fellowship
SFRH/BPD/26542/2006 (Portugal). The initial work of 
MBL was funded by MECD (Spain). MBL is also thankful 
to IMAFF (CSIC, Spain) and UBI (Portugal) for hospitality during the 
realisation of part of this work.

\end{document}